# Transverse rotation of the momentary field distribution and the orbital angular momentum of a light beam

A.Ya. Bekshaev

*I.I. Mechnikov National University, Dvoryanska 2, Odessa 65082, Ukraine*

*bekshaev@onu.edu.ua*

The transverse beam pattern, usually observed in experiment, is a result of averaging the optical-frequency oscillations of the electromagnetic field distributed over the beam cross section. An analytical criterion is derived that these oscillations are coupled with a sort of rotation around the beam axis. This criterion appears to be in direct relation with the usual definition of the beam orbital angular momentum.

Light beams carrying the mechanical orbital angular momentum (OAM) with respect to the propagation axis attract great attention due to their unique physical properties and promising applications [1–4]. The question of physical origin of the vortex properties of such beams is interesting and insistent [3–12]. In attempts to answer it, researchers have employed the momentary patterns of the light field behaviour during a cycle of oscillation [3, 4]. It was found that, in contrast to usual (non-vortex) beams where the evolution of the momentary transverse field distribution consists in periodical change of the field strength synchronously over the whole beam cross section, in the "classical" vortex beams (circular Laguerre-Gaussian modes) the momentary distribution of the electric field rotates with the field oscillation frequency (see Fig. 1a which illustrates the same behaviour as Fig. 10 of [3]).

However, this explanation of vortex properties of light beams does not seem satisfactory. First, the rate of the momentary pattern rotation of the sort depicted in Fig. 1a falls down for higher-order LG modes [4,10], whose vortex properties seem to be "stronger". Second, some classes of beams with OAM apparently show no rotation of the instantaneous field distribution, for example, rotating Gaussian beams with mismatching symmetries of the amplitude and phase profiles [7–9] whose momentary transverse patterns are presented in Fig. 1b. And, at last, this reasoning is based on the intuitive interpretation of the time evolution depicted in Fig. 1 which can be uncertain; it is essentially qualitative and needs a reliable quantitative foundation.

In this Letter, we intend (i) to derive a natural unambiguous analytical criterion for presence of the rotational component in the momentary beam pattern evolution and (ii) to show close relations of this criterion with usual definition of the beam OAM.

Consider a paraxial light beam propagating along axis $z$; transverse cross section is parameterized by coordinates $(x, y) \equiv \mathbf{r}$. We use a scalar model of the beam in which it can be characterized by the electric field component depending on the spatial coordinates and the time in accord with equation [3,4]

$$\mathcal{E}(\mathbf{r}, z, t) = u(\mathbf{r}, z) \exp\left[i(kz - \omega t)\right] \qquad (1)$$

where $\omega$ is the cyclic frequency of radiation, $k$ is its wave vector ($\omega = ck$ in a medium with the light velocity $c$) and $u(\mathbf{r}, z)$ is the slowly varying complex amplitude. Our further consideration will relate to a fixed cross section and, since it is always possible to choose the reference frame so that in this section $z = 0$, from now on, explicit indication of the longitudinal coordinate $z$ will be omitted. Eq. (1) is a complex representation of a harmonically oscillating field; the true instant field distribution at any time moment is determined by expression [13]

$$E(\mathbf{r}, t) = \text{Re}\left[\mathcal{E}(\mathbf{r}, t)\right]. \qquad (2)$$

Introducing the real module $A(\mathbf{r})$ and phase $\varphi(\mathbf{r})$ of the complex amplitude in accord with equation

$$u(\mathbf{r}) = A(\mathbf{r})\exp\left[ik\varphi(\mathbf{r})\right] \qquad (3)$$

the instant field (2) can be written in the form

$$E(\mathbf{r},t) = A(\mathbf{r})\cos\left[k\varphi(\mathbf{r}) - \omega t\right]. \qquad (4)$$

This equation describes behaviour of the momentary field distribution discussed in this Letter. In special cases of circular Laguerre-Gaussian (LG) modes, $A(\mathbf{r}) \equiv A(r)$, $k\varphi(\mathbf{r}) = l\phi$ where $r = \sqrt{x^2 + y^2}$, $\phi = \arctan(y/x)$ are the polar coordinates the transverse cross section, and $l$ is the integer mode index. Then

$$E(\mathbf{r},t) = A(r)\cos(l\phi - \omega t) \qquad (5)$$

so the rotation of the instant beam pattern is seen evidently (Fig. 1a). In many other typical situations the temporal behaviour of the field (4) represents a complicated oscillatory behaviour with local phases and amplitudes varying in different points, so one cannot tell "by eye" is there any rotational component in this motion or not.

We can try to find an analytical criterion for the presence of certain rotational component in the given oscillatory pattern. Let us note that for the case of "pure" rotation (5)

$$\frac{\partial E(\mathbf{r},t)}{\partial t} = -\frac{\omega}{l}\frac{\partial E(\mathbf{r},t)}{\partial \phi}. \qquad (6)$$

This establishes an exact correspondence between temporal and azimuthal derivatives of the instant field pattern. It could be expected that in more complicated cases, when Eq. (6) takes no place, certain correlation between the expressions standing in right and left parts of (6) can be a measure of presence of the sought rotational component.

So, the criterion that the instant field in the beam cross section shows a sort of rotational motion can be formulated as

$$\left\langle \frac{\partial E(\mathbf{r},t)}{\partial \phi} \frac{\partial E(\mathbf{r},t)}{\partial t} \right\rangle \propto \left\langle E_{\phi t} \right\rangle = \int \frac{\partial E(\mathbf{r},t)}{\partial \phi} \frac{\partial E(\mathbf{r},t)}{\partial t} (dr) dt \neq 0 \qquad (7)$$

where integration is performed over the whole cross section ($(dr) = dxdy$ is the element of area) and over the oscillation period. Integral $\left\langle E_{\phi t} \right\rangle$ in (7) is a usual measure of correlation between $\partial E / \partial \phi$ and $\partial E / \partial t$ and can be considered as a general "instant rotation criterion" (IRC). To avoid the dependence on the beam intensity, its expression (7) must be supplemented by due normalization procedure including mean values of the correlated quantities, i.e. integrals

$$\left\langle E_t^2 \right\rangle = \int \left( \frac{\partial E(\mathbf{r},t)}{\partial t} \right)^2 (dr) dt, \quad \left\langle E_\phi^2 \right\rangle = \int \left( \frac{\partial E(\mathbf{r},t)}{\partial \phi} \right)^2 (dr) dt \qquad (8)$$

which lead to the definition of a correlation coefficient (normalized IRC)

$$B = -\frac{\left\langle E_{\phi t} \right\rangle}{\left\langle E_t^2 \right\rangle^{1/2} \left\langle E_\phi^2 \right\rangle^{1/2}}. \qquad (9)$$

Negative sign in (9) makes the coefficient $B$ positive for the LG modes (5) when, in fact, perfect negative correlation takes place due to (6). For the prototype beam (5), the value $B = 1$ looks an adequate expression for the "perfect" rotational properties of the LG beam instant pattern.

Now consider expressions (7) – (9) with allowance for general definitions (3) and (4). Then

$$\frac{\partial E(\mathbf{r})}{\partial \phi} = \frac{\partial A(\mathbf{r})}{\partial \phi} \cos\left[k\varphi(\mathbf{r}) - \omega t\right] - A(\mathbf{r}) k \sin\left[k\varphi(\mathbf{r}) - \omega t\right] \frac{\partial \varphi(\mathbf{r})}{\partial \phi},$$

$$\frac{\partial E(\mathbf{r})}{\partial t} = A(\mathbf{r})\omega\sin\left[k\varphi(\mathbf{r}) - \omega t\right],$$

and calculations of (7), (8) are simplified because when integrating over the oscillation period squares of sines and cosines give the same value $\pi/\omega$ while their products give zeros. After substitution into (7) and (8) one obtains

$$\langle E_{\phi t}\rangle = -k\pi\int A^2(\mathbf{r})\frac{\partial \varphi(\mathbf{r})}{\partial \phi}(dr), \quad \langle E_t^2\rangle = \omega\pi\int A(\mathbf{r})^2(dr), \tag{10}$$

$$\langle E_\phi^2\rangle = \frac{\pi}{\omega}\int\left[\left(\frac{\partial A(\mathbf{r})}{\partial \phi}\right)^2 + k^2 A^2(\mathbf{r})\left(\frac{\partial \varphi(\mathbf{r})}{\partial \phi}\right)^2\right](dr). \tag{11}$$

Note that although the characteristics of instantaneous field pattern are now considered, their expressions (9) – (11) are represented through parameters of the complex amplitude (3), describing the usually observed time-averaged beam.

Look at the most important term – numerator of Eq. (9) represented by the first Eq. (10). Comparison with known formulas of [1,4,5,8] tells that it is proportional to the beam angular momentum (OAM); to be exact, if the electric field is given by (1),

$$\langle E_{\phi t}\rangle = -8\pi^2\omega\mathcal{L} \tag{12}$$

where $\mathcal{L}$ is the linear density of the beam OAM [1,4]. This is a central point of this Letter: the correlation coefficients (7), (9), which indicate presence of the rotational component in momentary variation of the transverse beam pattern, appear to be in the direct relation with the usual measure of the beam OAM. So, the vortex properties of the beam that are seen in very large, compared to the light oscillations, time scales, are directly linked to the momentary rotational transformations of the transverse beam pattern during the oscillation period. Any vortex behaviour of the time-averaged beam has its origin in this momentary rotation, despite that this rotation is a sort of "sunlight spot" and has no direct mechanical meaning [10,11] and even when the rotational component of oscillations is not intuitively obvious, as, for example, in case of rotating Gaussian beams (Fig. 1b).

Reciprocally, in any beam with OAM, the instantaneous transverse energy distribution $E(\mathbf{r},t)$ shows a sort of rotational motion in the sense of (7), (9).

The IRC (9) provides a quantitative expression of this conclusion but this is not the only possible choice. In particular, in accord with Eqs. (1) – (3), the second Eq. (10) leads to representation $\langle E_t^2 \rangle = 8\pi^2 k \Phi$ where $\Phi = (c/8\pi)\int A^2(\mathbf{r})(dr)$ is the total energy flow (power) of the beam [4,8]. Then, if we replace the second term in the denominator of (9) by the first one and multiply the whole expression by ω, with taking Eq. (12) into account we obtain another form of the IRC

$$\Lambda = -\omega \frac{\langle E_{\phi t} \rangle}{\langle E_t^2 \rangle} = \omega c \frac{\mathcal{L}}{\Phi} \tag{13}$$

which completely coincides with the normalized (specific) OAM [4] and is a measure of the beam OAM in units "$\hbar$ per photon" [4,5]. One more useful form of the IRC is formulated similarly:

$$\Omega = -\frac{\langle E_{\phi t} \rangle}{\omega \langle E_\phi^2 \rangle}. \tag{14}$$

Its application to the case of "pure" rotation (5), (6) shows that this amount has the meaning of effective angular velocity of the momentary pattern rotation measured in units of ω.

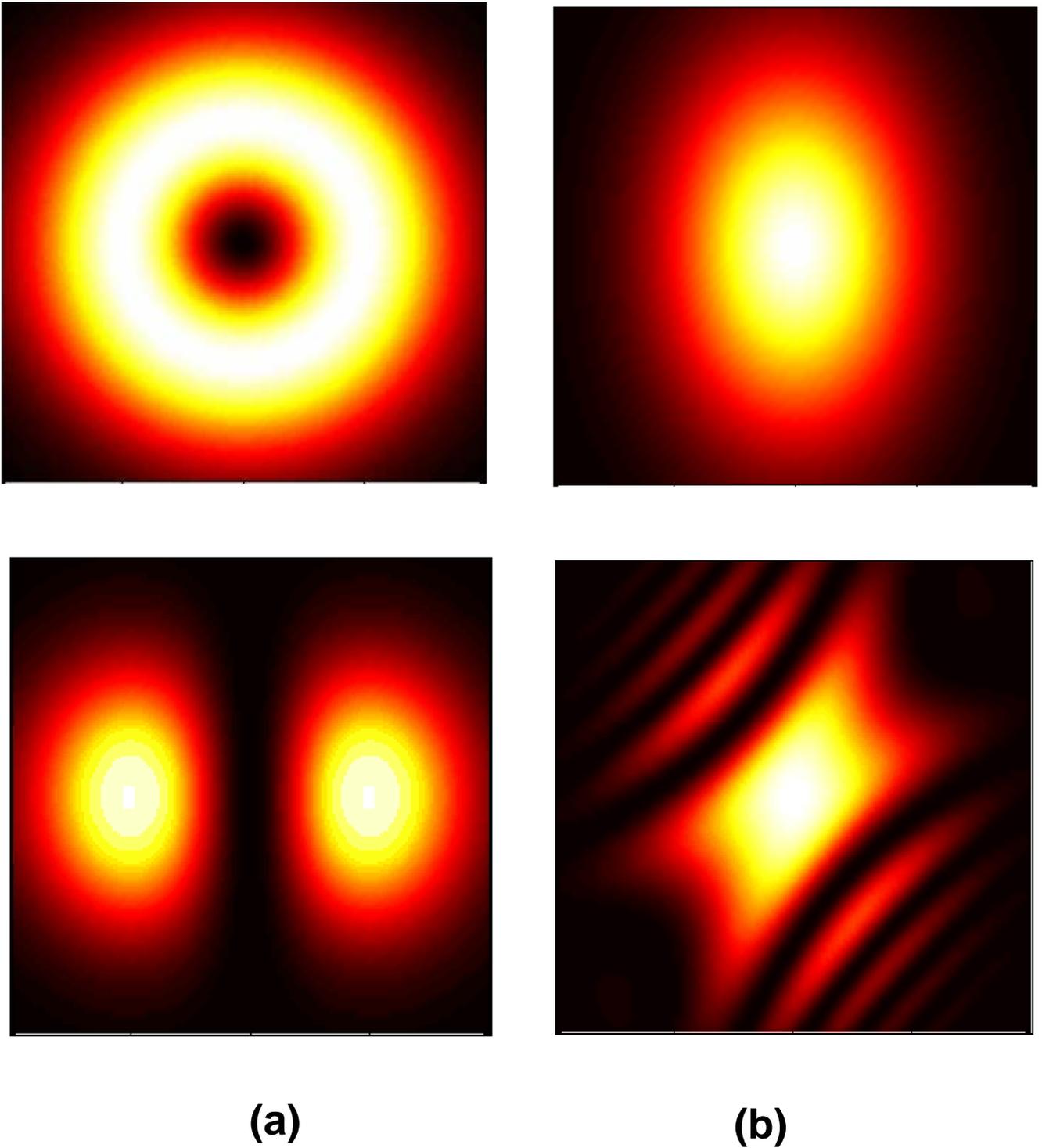

Fig. 1. Observable spot patterns (upper row) and evolution of the instantaneous electric field distribution (bottom row, click an image to play): (a) for LG mode (5) with $l=1$, $k=10^5$ cm$^{-1}$, $A(x,y) \propto \exp\left[-(x^2+y^2)/2b^2\right]$ with $b=0.1$ cm; (b) for the rotating Gaussian beam, complex amplitude is given by (3) where $A(x,y) \propto \exp(-x^2/2b_x^2 - y^2/2b_y^2)$, $\varphi(x,y) = 0.5(\alpha_{xx}x^2 + \alpha_{yy}y^2 + 2\alpha_{xy}xy)$ with $b_x = 0.1$ cm, $b_y = 0.15$ cm, $\alpha_{xx} = \alpha_{yy} = 2\cdot 10^{-3}$ cm$^{-1}$, $\alpha_{xy} = -3\cdot 10^{-3}$ cm$^{-1}$.